# The Dynamics of Creativity in Software Development[*]

Daniel Graziotin

Free University of Bozen-Bolzano

`daniel.graziotin@unibz.it`

**Abstract.** Software is primarily developed for people by people and human factors must be studied in all software engineering phases. Creativity is the source to improvise solutions to problems for dominating complex systems such as software development. However, there is a lack of knowledge in what creativity is in software development and what its dynamics are. This study describes the current state of the research plan towards a theory on creativity in software development. More specifically, it (1) states the motivation for studying creativity in software development under a multidisciplinary view; it (2) provides a first review of the literature identifying the shortcomings in the field; it (3) proposes a research design, which includes rarely employed methods in software engineering. To understand creativity in software development will provide a better knowledge of the software construction process and how individuals intellectually contribute to the creation of better, innovative products.

## 1      Introduction

As software is primarily developed by people for people, it is necessary to study human and social factors in all software engineering phases [6]. The advocates of Agile software movement emphasize the importance of people, to the point that 'People trump Process' [5]. Software development activities are perceived as creative and autonomous [18]. Software developers prefer to work on those development activities which are perceived as creative [14]. Creativity is crucial in software development as it is the source to solve complex problems and innovate [6].

Agile practitioners claim to rely on the creative talent of people to improvise solutions for complex software development problems [5, 15]. Traditional and formal processes rely instead on predictability and rationality in order to dominate systems. However, the environment in which software development happens is all but simple and predictable [10]. Too much change occurs while software is being developed and agility is required to adapt and respond to such changes [29]. Especially in small software organizations, environmental turbulence requires creativity to make sense of the changing environment [10].

Although the importance of creativity has been investigated in few publications [6, 14], the term appears to be misused in the literature. Little research has been done to explain what creativity is in software development and how this phenomenon arises. This might be explained by the insufficient clarity of the definitions of creativity and how to explore it [24]. The available definitions for creativity seem not to fit for many software development activities and artifacts. There is also a lack of an open-minded, multidisciplinary view when trying to explain complex objects such as creativity in a complex context like software development.

It has been recognized that software development is an intellectual and social activity [10] and it is carried out by cognitive processing activities [13, 17]. Creativity is cognitive and it is influenced by cognitive processes like emotions and mood [1, 8]. Much Psychology literature can be mined on this topic. However, there is the need to understand what creativity is in software development from a Software Engineering research perspective.

It is still necessary to exploit specialized disciplines like Psychology when studying the dynamics of creativity in software development, according to this author, because software development is a non-ordinary and complex system. For example, a client library software to access a remote API might have been automatically generated by another software, written during a "creative moment" after the observation of the patterns in repetitive programming activities. What and where would be creativity here? What caused such creative software development moment?

The purpose of this study is to explore the dynamics of creativity in software development activities and to explain what creativity is for such activities. By dynamics, this author means "the forces or properties that stimulate growth, development, or change of creativity in software development". There is the need to understand what creativity in software development activities is; what provokes creativity while developing software; how to influence the creativity of software engineers. The outcome of this author's PhD will be an evidence-based theory on creativity in software development. To understand creativity in software development will allow a better knowledge of the software construction process and how individuals intellectually contribute to the creation of better, innovative software products.

This paper summarizes the first four months of PhD of the author. Therefore, it concentrates on the literature review, the research questions and the proposed research methodology.

## 2      Literature Review

This section describes an ongoing literature review of creativity in Software Engineering research and Psychology. Although hundreds of definitions for creativity exist [24], most of them are related to the generation of products (ideas, solutions, artifacts) presenting (1) novelty and (2) usefulness [1, 8, 24]. Definitions tied to products, however, present issues in software development.

Creativity is believed to be beneficial and required in software development for decades. Brooks [2] considers the importance of creativity in *The Mythical Man-Month*.

Programming activity is usually fun because it enables creativity, although some activities such as bug fixing might not be creative at all [2, pp. 8-9]. Programmers are happy and optimist when they perceive their activities as creative.

Gu and Tong [14] report an exploratory research on creativity issues in software development. Students of a software architecture course implemented a software project. They filled a survey, which asked them to evaluate their work in terms of perceived creative time, perceived discipline-based time, and "other" time. From the results, the following hypotheses were formulated: (1) in software development, there is most creative work in the implementation phase and least creative work in the post-mortem analysis phase; (2) UML documentation promotes students to do more creative work in requirement specification and architecture design phases; (3) more creative work does neither accelerate nor decelerate development speed compared with discipline-based work; (4) developers prefer development phases including more creative work than discipline-based work.

Crawford et al. [6] report that previous research on creativity in software development primarily focused on requirement engineering and that the techniques to foster creativity are rarely investigated (brainstorming being a notable exception). They offer a linkage between basic types of creative thinking and requirements engineering. Then, the authors compare already proposed roles in a creative team with roles in eXtreme Programming. Finally, a linkage between eXtreme Programming activities and the creative process activities are given. No empirical research is reported.

Several proposals to foster creativity in software requirements exist – e.g., [20, 21, 23]. Although some authors conducted empirical research to evaluate techniques and tools, the evaluation is always in terms of the generated product, i.e., requirements. There is an explanation for this. In software engineering, requirements are arguably the artifacts, which resemble an idea more than anything else. Thus, the creativity of requirements can be easily understood and assessed as self-perceived creativity, or as defined in Psychology research.

Creativity has been studied in Psychology and Cognitive Science since more than 60 years ago and immediately acknowledged as being necessary for technology [26]. When dealing with creative performance, it is useful to distinguish between creative *product*, creative *process* [8], creative *person* and creative *press* (i.e., the relationship between humans and environment) [25]. In research, however, the outcomes of a creative performance often conceptualize the performance itself, in terms of novelty and value [8]. Additionally, affective states (mood, emotions, feelings) are believed to be "one of the most widely studied and least disputed predictors of creativity" [1].

In software development, explaining a creative performance by judging the creativity of a product might work for initial ideas and requirements, as well for the final software product. Alas, the creativity of intermediary products such as diagrams, architectures, and source-code seems difficult to be evaluated, if not pointless. However, it is currently unexplained how a creative performance produces such products and how this performance is structured.

Improvisation seems to play a role in the creativity of software developers and organizations in general. Improvisation is a "process of making sense of incoming working events and developing ad-hoc solutions", where "thinking and action seem to occur simultaneously" [4, pp. 369-371]. Despite improvisation is disregarded in the design of information systems and software processes in general, it is critical for software firms [3, 10]. Procedures and methods do not provide the implementation details of actions. Individuals interpret the methods according to human existence and experience [3]. Therefore, creativity is necessary to make sense of the ever-changing environment of software development [10]. Additionally, according to Ciborra, the act of improvisation is a mood [4, pp. 162-165] and the affective states enable the "mattering" of things. It has been argued that the ability to sense moods and emotions of software developers might be necessary for the success of an Information Technology company [9]. A linkage between improvisation, moods and creativity is likely and should be investigated in software development.

The research questions of this PhD are at an early stage. The following research questions have been proposed: (1) Is there a misconception of creativity in the context of software development? (2) What are the key components of creativity in software development at the individual, team, and organizational levels? (3) What is the relationship between creativity, moods, and improvisation in software development?

## 3   Proposed Research Methodology

Although Software Engineering is commonly treated as a scientific discipline, there have been criticisms on this approach. Software Engineering might be studied as a social discipline [4, 27]. According to this author, the truth lies somewhere in between and a pragmatic worldview is required. Pragmatism does not commit to any system of beliefs [7]. The best methods will be taken on a case basis, under a "whatever works at the time" philosophy (pp. 6-7 in [11]).

This author's PhD study is explorative in nature. The aim is to generate a theory from empirical evidence. There is the need to (1) define the steps of the theory building process, to (2) select a strategy to analyze the data, to (3) represent the theory in a meaningful way and to (4) choose research methods, in the order of importance to this study.

*Theory building process* Eisenhardt [12] proposes a sense-making process in theory building. The process of building theory is composed by eight main activities: the definition of the research questions, the selection of the cases, the crafting of the instruments and the protocols, the field entrance, the data analysis, the hypotheses shaping, the literature enfolding, and the closure reaching. These phases dictate the status of this PhD's research methodology.

*Data Analysis Methodology* As the aim of this PhD research is to generate insights, patterns, and theory from individuals' experiences, a strategy to gather and analyze the data like grounded theory [28] responds to such needs [19]. It is indicated to study human behavior in an iterative, explicit and systematic process [11]. Montoni and Rocha [22] employed it in Software Process Improvement studies [22]. They outline the phases of grounded theory research in six steps. First, the (1) context and the scope of the study are defined. Then, (2) data collection is defined and performed through dif-ferent methods – e.g., surveys, literature reviews, and (semi) structured interviews. After the data is collected, three different types of coding systems are required: open coding (3) – i.e., conceptualize and categorize the data; axial coding (4) – i.e., analysis of the relationships of the categories; selective coding (5) – i.e., the identification of the central category of the theory. The last phase of grounded theory research is to (6) apply audit techniques to ensure validity.

*Theory representation* Sjøberg et al. [16] suggest a framework for describing Software Engineering theories. A theory description should provide (1) its basic elements, or *constructs*, the (2) *propositions* describing the relationships of the constructs, the (3) *explanations* for the relationships, and the (4) *scope* of the theory. The framework provides the representations of the theory parts, using textual tables and a UML-like language with four archetype classes (Actor, Technology, Activity, and Software Systems).

*Research methods* As the *what* and the *why* of the topics under investigation are still not understood, qualitative studies will be conducted in the first part of the research activities. Unstructured interviews of software developers, managers, and creativity gurus will be carried out. Direct observations will extend the data from the interviews. A systematic mapping study on creativity in software development will enable further insights and enhance the validity of the theory that will emerge from the data.

Sjøberg et al. [16] suggest that a Software Engineering theory should be tested through empirical research. Therefore, Eisenhardt [12] process will be extended with a theory testing phase, employing cross-case analysis. Quantitative studies will be conducted. Surveys and experiments will provide quantitative evaluations of the relationships under study.

## 4      Conclusion

This paper proposed a research design to build a theory of creativity in software development. Since the author is at the beginning of the PhD study, the focus was given to the motivation of this study, on the multidisciplinary literature review, the research questions, and a plan for a research design and the framework for theory building.

## References

1. Baas, M. et al.: A meta-analysis of 25 years of mood-creativity research: hedonic tone, activation, or regulatory focus? Psychological Bulletin. 134, 6, 779–806 (2008).
2. Brooks, F.P.: The Mythical Man-Month. Addison-Wesley, Philippines (1975).
3. Ciborra, C.: Improvisation and information technology in organizations. ICIS 1996. p. 26 (1996).
4. Ciborra, C.: The Labyrinths of Information: Challenging the Wisdom of Systems. Oxford University Press, USA (2004).
5. Cockburn, A., Highsmith, J.: Agile software development, the people factor. Computer, IEEE. 34, 11, 131–133 (2001).
6. Crawford, B. et al.: Agile software teams must be creatives. 5th International Workshop on Co-operative and Human Aspects of Software Engineering. pp. 20–26 IEEE (2012).

7. Creswell, J.W.: Research design: qualitative, quantitative, and mixed method approaches. Sage Publications, Thousand Oaks, California (2009).
8. Davis, M.: Understanding the relationship between mood and creativity: A meta-analysis. Organizational Behavior and Human Decision Processes. 108, 1, 25–38 (2009).
9. Denning, P.J.: Moods. Communications of the ACM. 55, 12, 33 (2012).
10. Dyba, T.: Improvisation in small software organizations. Software, IEEE. September/October, 82–87 (2000).
11. Easterbrook, S. et al.: Selecting empirical methods for software engineering research. Guide to Advanced Empirical Software Engineering. 285–311 (2008).
12. Eisenhardt, K.: Building theories from case study research. Academy of management review. 14, 4, 532–550 (1989).
13. Fischer, G.: Cognitive View of Reuse and Redesign. IEEE Software. 4, 4, 60–72 (1987).
14. Gu, M., Tong, X.: Towards Hypotheses on Creativity in Software Development. 5th International Conference on Product Focused Software Process Improvement (PROFES). pp. 47–61 Springer Verlag (2004).
15. Highsmith, J., Cockburn, A.: Agile software development: the business of innovation. Computer, IEEE. 34, 9, 120–127 (2001).
16. Jorgensen, M., Sjoberg, D.: Generalization and theory-building in software engineering research. 1, 1–7 (2004).
17. Khan, I.A. et al.: Do moods affect programmers' debug performance? Cognition, Technology & Work. 13, 4, 245–258 (2010).
18. Knobelsdorf, M., Romeike, R.: Creativity as a pathway to computer science. ACM SIGCSE Bulletin. 40, 3, 286 (2008).
19. Langley, A.: Strategies for Theorizing from Process Data. Academy of Management Review. 24, 4, 691 (1999).
20. Maiden, N. et al.: Provoking Creativity: Imagine What Your Requirements Could Be Like. IEEE Software. 21, 05, 68–75 (2004).
21. Maiden, N., Hollis, B.: Extending Agile Processes with Creativity Techniques. IEEE Software. 1–1 (2012).
22. Montoni, M.A., Rocha, A.R.: Applying Grounded Theory to Understand Software Process Improvement Implementation. 7th International Conference on the Quality of Information and Communications Technology. pp. 25–34 IEEE (2010).
23. Obrenovic, Z. et al.: Stimulating creativity through opportunistic software development. Software, IEEE. November / December, 64–70 (2008).
24. Piffer, D.: Can creativity be measured? An attempt to clarify the notion of creativity and general directions for future research. Thinking Skills and Creativity. 7, 3, 258–264 (2012).
25. Rhodes, M.: An Analysis of Creativity. The Phi Delta Kappan. 42, 7, 305–310 (1961).
26. Simonton, D.: Creativity: Cognitive, personal, developmental, and social aspects. American Psychologist. 55, 1, 151–158 (2000).
27. Sjøberg, D. et al.: Building theories in software engineering. Guide to Advanced Empirical Software Engineering. 1, 1, 312–336 (2008).
28. Strauss, A.L., Corbin, J.M.: Basics of Qualitative Research: Techniques and Procedures for Developing Grounded Theory. Sage Publications, London (2008).
29. Williams, L., Cockburn, A.: Agile Software Development: It's about Feedback and Change. Computer, IEEE. June, 39–43 (2003).